\documentclass[%
reprint,
 amsmath,amssymb,
 aps,
]{revtex4-1}

\usepackage{graphicx}
\usepackage{dcolumn}
\usepackage{bm}
\usepackage{braket}
\usepackage{hyperref}
\usepackage{cleveref}
\DeclareMathOperator{\Tr}{Tr}

\begin{document}

\preprint{APS/123-QED}

\title{Thermodynamics of a Higher-Order Topological Insulator}

\author{R. Arouca$^{1,2}$}\thanks{r.aroucadealbuquerque@uu.nl}
\author{S. N. Kempkes$^{1}$}\thanks{S.N.Kempkes@uu.nl}\author{C. Morais Smith$^{1}$}\thanks{C.deMoraisSmith@uu.nl}
\affiliation{$^1$Institute for Theoretical Physics, Center for Extreme Matter and Emergent Phenomena, Utrecht University, Princetonplein 5, 3584 CC Utrecht, The Netherlands.}
\affiliation{$^2$Instituto de F\' isica, Universidade Federal do Rio de Janeiro, C.P. 68528, Rio de Janeiro, RJ, 21941-972, Brazil.}%

\date{\today}

\begin{abstract}

	We investigate the order of the topological quantum phase transition in a two dimensional quadrupolar topological insulator within a thermodynamic approach. Using numerical methods, we separate the bulk, edge and corner contributions to the grand potential and detect different phase transitions in the topological phase diagram. The transitions from the quadrupolar to the trivial or to the dipolar phases are well captured by the thermodynamic potential. On the other hand, we have to resort to a grand potential based on the Wannier bands to describe the transition from the trivial to the dipolar phase. The critical exponents and the order of the phase transitions are determined and discussed in the light of the Josephson hyperscaling relation.

\end{abstract}

\maketitle


\section{Introduction}
\label{sec_intro}

	Topological states of matter have attracted a great deal of attention during the last decade. In addition to the usual discrete symmetries that compose the tenfold way~\cite{altland1997nonstandard,schnyder2008classification, kitaev2009periodic}, the periodic table of topological materials was extended by the inclusion of crystalline symmetries~\cite{fu2011topological,slager2013space}, which led to the prediction of new kinds of topological phases. A typical example is the recent proposal of higher-order topological insulators (HOTIs)~\cite{benalcazar2017_Science,*benalcazar17_PRB}.

	HOTIs, or more generally higher-order topological phases, are such that the symmetry protected modes (SPMs) associated with the boundary between the topological phase and the vacuum occur not in the system with codimension 1, but with codimension 2 or higher. This means that for two-dimensional (2D) systems, these modes will appear at the 0D corners. Similarly, for 3D systems they do not arise at the 2D surface but at the 1D hinges or at the 0D corners of the system. The existence of HOTIs was experimentally verified in a variety of systems, such as bismuth~\cite{schindler2018higher}, electrical circuits~\cite{imhof2018topolectrical,serra2019observation}, acoustic~\cite{ni2019observation,xue2019acoustic} and electronic systems~\cite{kempkes2019robust}, and new types of HOTIs and higher-order topological superconductors were recently proposed~\cite{matsugatani2018connecting, ezawa2018higher, cualuguaru2019higher, agarwala2019higher}.

	A very interesting question posed by this kind of topological phase is the fate of the bulk-boundary correspondence~\cite{trifunovic2019higher, khalaf2019boundary}.~A thermodynamic analysis of topological quantum phase transitions (TQPT)~\cite{quelle2016thermodynamic, kempkes2016universalities,van2018thermodynamic,cats2018staircase} inspired by Hill's thermodynamics~\cite{hill1994thermodynamics,chamberlin99, chamberlin2000mean,chamberlin2015big,latella2015thermodynamics} has been successful in describing the critical behaviour of several models and recently this method was used to devise topological heat machines~\cite{yunt2019topological}.~Here, we extend this thermodynamical approach to describe HOTIs, i.e., to deal with systems that have both edge and corner modes. For this, we investigate the model proposed in Ref.~\cite{benalcazar2017_Science,*benalcazar17_PRB}, which is a 2D SSH model, with broken symmetry between the $x$ and $y$ directions and different intra- and inter-cell hopping parameters. In addition, a $\pi$-flux pierces the plaquettes, such that some of the hopping parameters are negative. This model exhibits a rich phase diagram, with phases characterised by the polarisation: depending on the ratio between the different hopping parameters, the system is in a quadrupolar, a dipolar or a trivial phase. To describe the phase transitions, we calculate the correlation functions for the bulk, edge and corner contributions, and look for divergences or jumps in their derivatives, which signal a phase transition. 

	For the transition from the trivial to the quadrupolar state, we find that the bulk exhibits a third-order, the edge a second-order, and the corner a first-order phase transition. The transition from the dipolar to the quadrupolar phase is also visible in the thermodynamic response at the edge and corner contributions, although it is invisible for the bulk. Finally, the transition from the trivial to dipolar phase cannot be captured by the thermodynamic potential description, since there is no band gap closing at the transition. To characterise the topology of these phases, the Wannier spectrum was investigated~\cite{benalcazar2017_Science,*benalcazar17_PRB}. The gap in the spectrum of the Wannier centers closes for all phase transitions, and therefore we extracted an effective grand potential from the Wannier spectrum. We found that the Wannier grand potential exhibits discontinuities at all phase transitions, fully characterising the critical behaviour of the system. The critical exponents were extracted by analysing the functional dependence of the closing of either the band gap or the Wannier gap, and the order of the phase transitions were determined using the Josephson hyperscaling relation. An excellent agreement is found with the order obtained by analysing the discontinuities in the thermodynamic or the Wannier grand potential. 
	
	The outline of the paper is as follows: In Section \ref{sec_quad}, we present the quadrupolar topological insulator model proposed in Ref.~\cite{benalcazar2017_Science,*benalcazar17_PRB}, focusing specifically on the different phases of the phase diagram. Next, in Section \ref{sec_hill}, we extend the thermodynamical description of topological materials to deal with systems composed of bulk, edge and corner contributions, and explain how one can devise a subtraction scheme to identify the critical behavior of each part. In Section \ref{sec_quad_phase transition}, we apply this approach to investigate the quadrupolar to trivial phase transition and show how the order of the phase transition can be understood by determining the critical exponents and using the Josephson's hyperscaling relation. In Section \ref{sec_dip_phase transition}, we perform the same analysis to other phase transitions in the phase diagram and show that the procedure does not apply due to the absence of a gap closing in some of the topological phase transitions. In Section \ref{sec_wann_free}, we show that a grand potential calculated from the Wilson loop spectra, that we name Wannier grand potential, is sensitive to all phase transitions, but with critical exponents of a system with smaller dimension, which we identify to be the edge of the original lattice.

\section{Quadrupolar topological insulator}\label{sec_quad}

	The existence of HOTIs was proposed based on the modern theory of polarisation~\cite{benalcazar2017_Science,*benalcazar17_PRB,resta1994modern,spaldin2012beginner,neupert2018lecture}, which associates the projected position operator to the Wilson loop and a topological index. In these systems, there are SPMs that are localised not at the entire surface of the material, but on parts of this surface. These modes are analogous to multipoles in classical electrodynamics. To realise a quadrupolar topological insulator in a 2D lattice, such that the protected modes are localised at the corners of the lattice, Benalcazar \text{et al.}~\cite{benalcazar2017_Science,*benalcazar17_PRB} proposed a model which, in presence of $C_4$ symmetry, has a vanishing dipolar order and a nonzero charge localised in the corner due to the bulk quadrupolar moment \cite{khalaf2019boundary}.

\begin{figure}[!ht]
		\includegraphics[width=\linewidth]{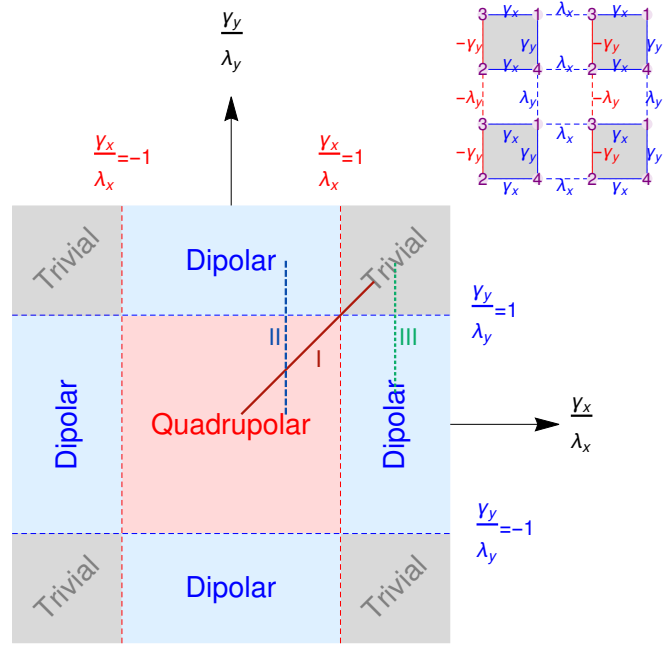}
		\caption{(Color online) Phase diagram of Hamiltonian~\eqref{eq_ham} showing the topological phases. For $\left|\gamma_x/\lambda_x\right|\leq 1$ and $\left|\gamma_y/\lambda_y\right|\leq 1$ the system is in the quadrupolar phase. The dipolar phases are the ones where only one of the ratios $\left|\gamma_x/\lambda_x\right|$ or $\left|\gamma_y/\lambda_y\right|$ is smaller than 1; when both are larger than 1, the system is in the trivial state. We will consider in the following sections the transitions between: the trivial and quadrupolar state (path I in red), the dipolar to quadrupolar state (path II in blue), and the trivial to dipolar state (path III in green). Inset: schematic image of the lattice for Hamiltonian~\eqref{eq_ham}. The intercell hopping parameters $\lambda$ are represented by dashed lines, while the intracell hopping parameters $\gamma$ are represented by solid lines. Negative values of hopping, created by a $\pi$-flux, are represented by red links.}
		\label{fig_phase}
	\end{figure}

	The Hamiltonian of this model reads  
	\begin{align}
	\begin{split}
		H=\sum\limits_{\mathbf{R}}&\gamma_x\left[c^\dagger_{\mathbf{R},1}c_{\mathbf{R},3}+c^\dagger_{\mathbf{R},2}c_{\mathbf{R},4}+h.c.\right]\\
		+&\gamma_y\left[c^\dagger_{\mathbf{R},1}c_{\mathbf{R},4}-c^\dagger_{\mathbf{R},2}c_{\mathbf{R},3}+h.c.\right]\\
		+&\lambda_x\left[c^\dagger_{\mathbf{R},1}c_{\mathbf{R}+\mathbf{x},3}+c^\dagger_{\mathbf{R}+\mathbf{x},2}c_{\mathbf{R},4}+h.c.\right]\\
		+&\lambda_y\left[c^\dagger_{\mathbf{R},1}c_{\mathbf{R}+\mathbf{y},4}-c^\dagger_{\mathbf{R},3}c_{\mathbf{R}+\mathbf{y},2}+h.c.\right],
	\end{split}
		\label{eq_ham}
	\end{align}
	where $c^{\left(\dagger\right)}_{\mathbf{R},i}$ denotes the destruction (creation) operator at site $i$ in the unit cell labelled by $\mathbf{R}$ (see sketch in the inset of Fig.~\ref{fig_phase}), and $\gamma_{x,y}$ ($\lambda_{x,y}$) are the intracell (intercell) hopping parameters in the $x$ or $y$ directions, respectively. Notice the negative hopping terms in Eq.~\eqref{eq_ham}, which correspond to a $\pi$-flux (gauge field) piercing the plaquettes. The model exhibits a quadrupolar phase transition when the hopping parameters are $\gamma_x=\gamma_y=\gamma$, $\lambda_x=\lambda_y=1$ and $\left|\gamma\right|<1$, such that the different corners are related by $C_4$ symmetry and the $\pi$-flux in the hopping cancels the edge currents. 

	The phase diagram of this system is shown in Fig.~\ref{fig_phase}~\cite{benalcazar2017_Science, *benalcazar17_PRB}. The topological index that characterises the topological phase is $\mathbf{p}=\left(p_x^{\zeta_y^{-}},p_y^{\zeta_x^{-}}\right)$, where $p_{x/y}^{\zeta_{y/x}^{-}}$ is the $x/y$ polarisation of the Wannier band $\zeta_{x/y}\left(k_y/k_x\right)$ localised below the gap. This polarisation is quantized in units of $1/2$ and is defined$ \mod 1$, such that it is either $0$ or $1/2$, characterizing a $\mathbb{Z}_2\times\mathbb{Z}_2$ invariant. The trivial phase has $\mathbf{p}=\left(0,0\right)$, the dipolar phase has $\mathbf{p}=\left(1/2,0\right)$ or $\mathbf{p}=\left(0,1/2\right)$ and the quadrupolar phase has $\mathbf{p}=\left(1/2,1/2\right)$. As the polarisation is calculated from the Wannier bands related to the bulk spectrum, the appearance of zero modes in the corner is related to a gap closing in the bulk, showing a new kind of bulk-boundary correspondence~\cite{trifunovic2019higher, khalaf2019boundary}.

\section{Thermodynamics of the quadrupolar phase transition}\label{sec_hill}

	TQPTs are QPTs and therefore one should in principle be able to describe their thermodynamic behaviour. However, the Ginzburg-Landau scheme fails to characterise them because they do not have a local order parameter. In addition, taking the thermodynamic limit is also a problem because one is interested in the edges of the system and these do not contribute in this limit anymore. Nevertheless, a scaling analysis can be used \cite{chen2016scaling, chen2018weakly, chen2019universality, molignini2019generating, molignini2020unifying, continentino2017quantum, CONTINENTINO2017A1, griffith2018casimir, rufo2019multicritical, wang2018decoding} and, particularly, it was shown in previous studies \cite{quelle2016thermodynamic, kempkes2016universalities,van2018thermodynamic,cats2018staircase} that a generalized Ehrenfest's classification \cite{Pippard, jaeger} still holds for describing topological phase transitions  and allows to characterise them by the order of discontinuity or divergence of the derivative of the grand potential.

	The bulk-boundary correspondence relates the nontrivial topology of the bulk, which is associated with a winding number, to the presence of SPMs at the boundary. In this way, the phase transition in the bulk (signalled by the gap closing) is connected to the one at the boundaries (signalled by the appearance/disappearance of the SPMs). However, the bulk and the boundary scale differently with the size of the system. Therefore, a thermodynamic description of TQPTs is more subtle. This leads to a non-extensive grand potential, as only the part of the grand potential related to the bulk scales with the number of particles in the system.

	The solution to this conundrum is provided by an approach inspired by Hill thermodynamics~\cite{hill1994thermodynamics,quelle2016thermodynamic, kempkes2016universalities, van2018thermodynamic, cats2018staircase,yunt2019topological,chamberlin99, chamberlin2000mean,chamberlin2015big,latella2015thermodynamics, Li2014, hill2001different, bedeaux2018}. Within this method, the system is divided into many subsets and the thermodynamic identity is derived without the assumption of extensivity of the energy. Hence, one can describe a finite size system, for which the bulk, the edge and the corner contributions scale differently with the system size. For the 2D system under investigation, we extended this analysis to deal with systems for which the bulk scales with $L^2$, the edge with $L^{1}$ and the corner with $L^0$. The grand potential then can be written as
	\begin{align}
	\begin{split}
		\Omega&=\Omega_{\text{bulk}}+\Omega_{\text{edge}}+\Omega_{\text{corner}}\\
		&=\omega_{\text{bulk}}L^2+\omega_{\text{edge}}L+\omega_{\text{corner}},
	\end{split}
		\label{eq_Omega_bec}
	\end{align}
	where the $\omega$'s are the intensive contributions for each component of the system. The scaling of the different $\omega$ shown in the Appendix confirms that indeed they are intensive quantities.

	\begin{figure}[!ht]
		\centering
		\includegraphics[width=\linewidth]{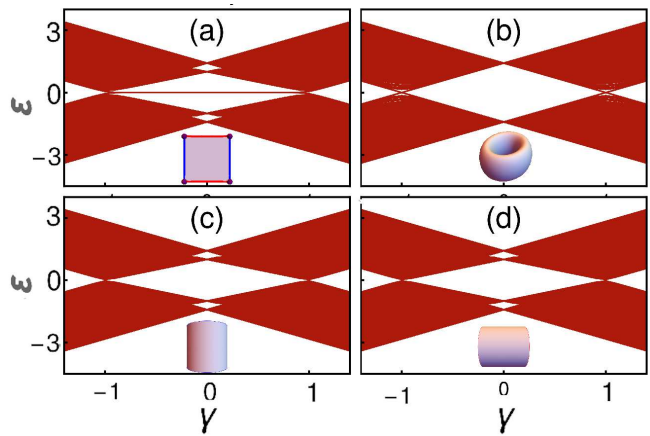}
		\caption{(Color online) Spectrum of the Hamiltonian~\eqref{eq_ham} for: (a) open, (b) fully-periodic, (c) $x$-periodic and (d) $y$-periodic boundary conditions. We fix $\lambda_x=\lambda_y=1$ and vary the intracell hopping parameters as $\gamma_x=\gamma_y=\gamma$ (path I in Fig.~\ref{fig_phase}) for a lattice with 40x40 unit cells. The TQPT occurs for $\left|\gamma\right|=1$, at which there is a gap closing and the emergence/disappearance of zero-energy modes. The zero modes are present only with open boundary conditions, confirming that indeed these modes are due to corner states. The similarity between the spectrum of the $x$-periodic and $y$-periodic is due to the $C_4$ symmetry of Hamiltonian~\eqref{eq_ham} for the chosen parameter values.}
		\label{fig_spectrum}
	\end{figure}

	For the Hamiltonian~\eqref{eq_ham}, we can write $\Omega$ as
	\begin{equation}
		\Omega=-\frac{1}{\beta} \ln\left\{\Tr\left[e^{-\beta\left(\hat{H}-\mu \hat{N}\right)}\right]\right\},
		\label{eq_gran_pot}
	\end{equation}
with $\hat{N}$ denoting the number operator, $\mu$ the chemical potential and $\beta=1/k_BT$. One can then perform the calculation for different system sizes and separate the components in Eq.~\eqref{eq_Omega_bec}, by verifying how $\Omega$ scales with the system size~\cite{quelle2016thermodynamic,kempkes2016universalities}.

\begin{figure}[!ht]
		\centering
		\includegraphics[width=\linewidth]{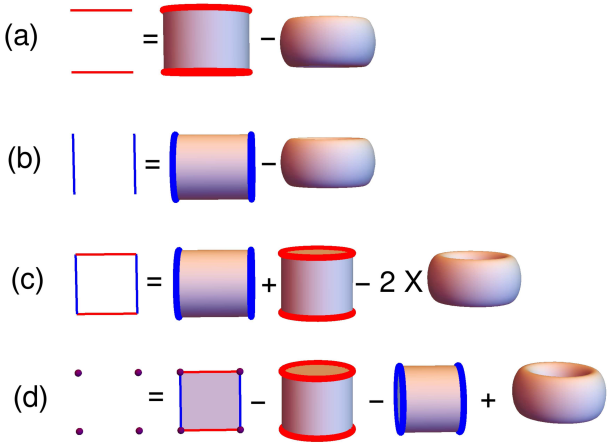}
		\caption{(Color online) Representation of the subtraction schemes used to obtain the boundary contributions to the grand potential. (a) Scheme to obtain the $y$-edge contribution from the $x$-periodic and fully-periodic boundary condition systems, (b) similarly for the $x$-edge contribution. (c) Scheme to obtain the complete edge contribution from the $y$-periodic, $x$-periodic and fully-periodic boundary condition systems and (d) scheme to obtain the corner contribution from the open, $x$-periodic, $y$-periodic and fully-periodic boundary condition systems.} 
		\label{fig_sum_corner}
	\end{figure}

	However, there is another scheme to obtain these contributions using a more geometrical approach~\cite{cats2018staircase}. Depending on the boundary conditions of the system, some of these terms are not present, as suggested from the different spectra in Fig.~\ref{fig_spectrum}. For instance, if we use periodic boundary conditions, there is no boundary between the system and a trivial vacuum. Hence, there is neither a corner nor an edge contribution to the grand potential. Similarly, if we use periodic boundary conditions only along the $x$ or $y$ direction, the system will have no corner contribution, but just one of the edge contributions. Therefore, we can isolate the different terms in Eq.~\eqref{eq_Omega_bec} by subtracting the grand potential of systems with different boundary conditions. We represent pictorially the process of subtraction in Fig.~\ref{fig_sum_corner}.

\subsection{Correlation functions in thermodynamics}\label{subsec_corr}

	Instead of dealing with the grand potential, we will consider its derivative. The derivative is related to correlation functions, which will allow us to identify the phase transition. As a matter of fact, if the control parameter of the phase transition is $\eta$, the derivative of $\Omega$ with respect to $\eta$ in Eq.~\eqref{eq_gran_pot} is
	\begin{align}
	\begin{split}
		\frac{\partial \Omega}{\partial \eta}=&\frac{1}{\mathcal{Z}}\Tr\left\{\left[\frac{\partial \hat{H}}{\partial \eta}-\frac{\partial \mu}{\partial \eta}\hat{N}\right]e^{-\beta\left(\hat{H}-\mu \hat{N}\right)}\right\}=\\
		=&\Braket{\frac{\partial \hat{H}}{\partial \eta}-\frac{\partial \mu}{\partial \eta}\hat{N}},
	\end{split}
		\label{eq_Omega_x}
	\end{align}
and this is equal to a correlation function for a Hamiltonian in second quantization. Hence, the discontinuity in $\Omega$ is encoded in this correlation function~\cite{cats2018staircase}.

    From the set of eigenstates, we can obtain the correlation functions using a matrix $\mathbb{S}$ that relates a vector composed of the eigenstates $\mathbf{\Psi}$ and a vector composed of the annihilation operators $\mathbf{C}$
	\begin{equation}
	\begin{cases}
		\mathbf{\Psi}=\mathbb{S}^{-1}\mathbf{C}\\
		 \mathbf{C}=\mathbb{S} \mathbf{\Psi}
	\end{cases} 
	\end{equation}
and we can write a correlation function for the $c$ operators as 
	\begin{equation}
		\Braket{c^\dagger_{\rho}c_{\sigma}}\left(T,\mu\right)=\sum\limits_{m,n}S^{*}_{\rho,m}S_{\sigma,n}\Braket{\psi^\dagger_m\psi_n}\left(T,\mu\right),
	\end{equation}
where $\rho$ and $\sigma$ are labels for the $c$ operators and $m$ and $n$ for the energy states. As $\psi^\dagger$ is the creation operator of the eigenstates of the Hamiltonian, the correlation function $\Braket{\psi^\dagger_m\psi_n}\left(T,\mu\right)$ will be simply the occupation of each energy level, given by the Fermi-Dirac distribution multiplied by $\delta_{m,n}$
	\begin{equation}
		\Braket{c^\dagger_{\rho}c_{\sigma}}\left(T,\mu\right)=\sum\limits_{m}S_{\rho,m}^*S_{\sigma,m}\frac{1}{e^{\beta\left(\varepsilon_{m}-\mu\right)}+1}.
		\label{eq_corr_finite}
	\end{equation}
	
	Hence, by knowing the spectrum of the system, one can calculate the correlation functions. In particular, for the quantum phase transition under consideration, we evaluate the correlation function for $T=0$ and $\mu=0$. In this case, the Fermi-Dirac distribution is a step function, which vanishes for $\varepsilon>\varepsilon_{F}=0$ and Eq~\eqref{eq_corr_finite} reduces to
	\begin{equation}
		\Braket{c^\dagger_{\rho}c_{\sigma}}\left(T=0,\mu=0\right)=\sum\limits_{\varepsilon_m\leq0}S_{\rho,m}^*S_{\sigma,m}.
		\label{eq_corr_T0_mu0}
	\end{equation}
	
	In this way, the correlation functions can be obtained by diagonalizing $\hat{H}$. Notice that for different boundary conditions, the spectrum of the system and the wavefunctions can be different, which ultimately lead to different correlation functions.

\subsection{Correlation functions for the quadrupolar transition}\label{subsec_corr_quad}

	To describe the thermodynamic response of a system divided in bulk, edge and corner, we take the derivatives of each term in Eq.~\eqref{eq_Omega_bec}. For the Hamiltonian~\eqref{eq_ham}, we write the derivative of $\Omega$ with respect to $\gamma_x=\gamma_y=\gamma$ given by Eq.~\eqref{eq_Omega_x} as
	\begin{align}
	\begin{split}
		\frac{\partial \Omega}{\partial \gamma}&=\sum\limits_{\mathbf{R}}\Braket{c^\dagger_{\mathbf{R},1}\left(c_{\mathbf{R},3}+c_{\mathbf{R},4}\right)+h.c.}\\
			&+\Braket{c^\dagger_{\mathbf{R},2}\left(c_{\mathbf{R},4}-c_{\mathbf{R},3}\right)+h.c.}\\
		&=\Braket{C^\dagger C},
	\end{split}
		\label{eq_Omega_dgamma}
	\end{align}
where in the last line we introduced the notation $\Braket{C^\dagger C}$ for the whole correlation function.

	To obtain the terms of Eq.~\eqref{eq_Omega_bec}, we identify the different contributions for each boundary condition following the scheme of Fig.~\ref{fig_sum_corner}. The scheme follows from the scaling analysis in Eq.~\eqref{eq_Omega_bec}. For a fully periodic system, the correlation function is the same for each unit cell. Hence, the total correlation function is calulated by multiplying the one for an individual unit cell with the number of unit cells. This contribution scales with $L^d$, where $d$ is the dimension. When considering semi-periodic boundary conditions, the same argument holds for a super cell along the open boundary and the correlation function scales with $L^{d-1}$. Since the spectrum is the same for the periodic and semi-periodic boundary conditions, apart from the boundary modes, we obtain the contributon of these boundary modes when subtracting the two contributions. This means we can assume that the contributions coming from this system contains only $\Omega_\text{bulk}$, such that
	\begin{equation}
		\frac{\partial \Omega_\text{bulk}}{\partial \gamma}
		=\Braket{C^\dagger C}^{\text{per}}=L^2\frac{\partial \omega_\text{bulk}}{\partial \gamma},
		\label{eq_omega_bulk_dgamma}
	\end{equation}
	where the upper index "per" denotes that the correlation function was calculated using periodic boundary conditions. As the system is periodic, the above correlation function does not depend on the position and the sum will be just $L^2$ times the correlation function for some point that gives us $\omega_\text{bulk}$.

	Then, if we subtract the correlation functions associated with the fully-periodic boundary conditions system from the one with open boundary conditions,
	\begin{align}
	\begin{split}
		\frac{\partial \Omega}{\partial \gamma}-\frac{\partial \Omega_\text{bulk}}{\partial \gamma}=&\Braket{C^\dagger C}^{\text{open}}-\Braket{C^\dagger C}^{\text{per}}\\
		=&\frac{\partial \Omega_\text{edge}}{\partial \gamma}+\frac{\partial \Omega_\text{corner}}{\partial \gamma},
	\end{split}
	\end{align}
	we obtain only the edge and corner contributions to the grand potential.
	
	Next, we assume that the $x$-periodic system has only the $y$-edge and the bulk contributions,
	\begin{align}
	\begin{split}
		&\frac{\partial \Omega_\text{$y$-edge}}{\partial \gamma}+\frac{\partial \Omega_\text{bulk}}{\partial \gamma}=\Braket{C^\dagger C}^{\text{x-per}},
	\end{split}
		\label{eq_Omega_y_dgamma}
	\end{align}
and similarly for the $y$-periodic system,
	\begin{align}
	\begin{split}
		&\frac{\partial \Omega_\text{$x$-edge}}{\partial \gamma}+\frac{\partial \Omega_\text{bulk}}{\partial \gamma}=\Braket{C^\dagger C}^{\text{y-per}}.
	\end{split}
		\label{eq_Omega_x_dgamma}
	\end{align}
	Therefore, the edge term corresponding to the sum of $\Omega_\text{$x$-edge}+\Omega_\text{$y$-edge}$ will be
	\begin{align}
	\begin{split}
		\hspace{-0.3cm}\frac{\partial \Omega_\text{edge}}{\partial \gamma}&=\frac{\partial \Omega_\text{$y$-edge}}{\partial \gamma}+\frac{\partial \Omega_\text{$x$-edge}}{\partial \gamma}\\
		&=\Braket{C^\dagger C}^{\text{y-per}}\hspace{-0.4cm}+\Braket{C^\dagger C}^{\text{x-per}}\hspace{-0.4cm}-2\Braket{C^\dagger C}^{\text{per}}\\
		&=L\frac{\partial \omega_\text{edge}}{\partial \gamma}.
	\end{split}
		\label{eq_omega_edge_dgamma}
	\end{align}
	Finally, the corner contribution can be obtained when we subtract the edge and bulk parts,
	\begin{align}
	\begin{split}
		&\frac{\partial \Omega}{\partial \gamma}-\frac{\partial \Omega_\text{bulk}}{\partial \gamma}-\frac{\partial \Omega_\text{edge}}{\partial \gamma}=\\
	&=\Braket{C^\dagger C}^{\text{open}}\hspace{-0.5cm}-\Braket{C^\dagger C}^{\text{y-per}}\hspace{-0.4cm}-\Braket{C^\dagger C}^{\text{x-per}}\hspace{-0.5cm}+\Braket{C^\dagger C}^{\text{per}}\\
	&=\frac{\partial \Omega_{\text{corner}}}{\partial \gamma}=\frac{\partial \omega_{\text{corner}}}{\partial \gamma}
	\end{split}
		\label{eq_omega_corner_dgamma}
	\end{align}
	Calculating then the correlation functions from Eq.~\eqref{eq_Omega_dgamma} for the open, fully-periodic, $x$-periodic and $y$-periodic boundary functions, we can obtain the different contributions in Eq.~\eqref{eq_Omega_bec}.
	
    To sum up, using Eq.~\eqref{eq_corr_T0_mu0} with $\rho$ and $\sigma$ taking the lattice indices of Eq.~\eqref{eq_Omega_dgamma}, one can obtain the derivatives with respect to $\gamma$ of the different contributions to the grand potential in Eq.~\eqref{eq_Omega_bec}, using the correlations functions with the boundary conditions described in Eqs.~\eqref{eq_omega_bulk_dgamma},~\eqref{eq_omega_edge_dgamma} and~\eqref{eq_omega_corner_dgamma}, which follow the subtraction scheme of Fig.~\ref{fig_sum_corner}.
	\begin{figure}[!ht]
		\centering
		\includegraphics[width=\linewidth]{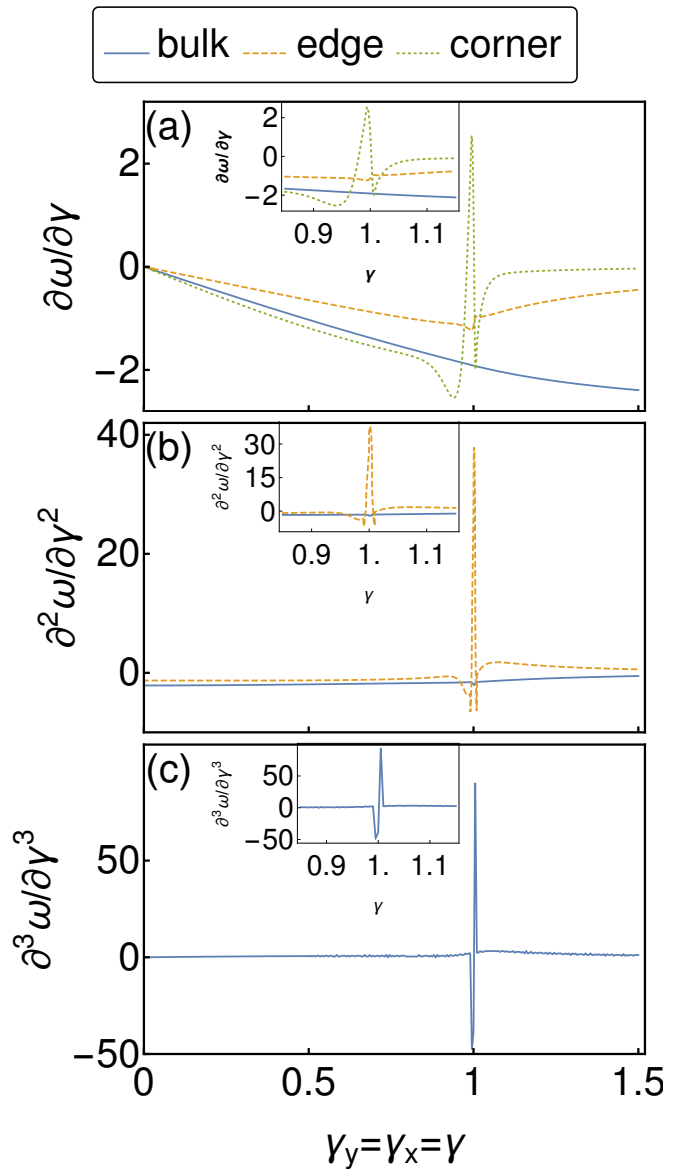}
		\caption{(Color online) (a) First, (b) second and (c) third derivatives of the grand potential with respect to $\gamma_x=\gamma_y=\gamma$ (path I in Fig.~\ref{fig_phase}) for the bulk (blue), edge (yellow) and corner (green) contributions. We fix $\lambda_x=\lambda_y=1$. The lattices have $40\times40$ unit cells. The inset shows a zoom in on the region of the phase transition.}
		\label{fig_corr_g}
	\end{figure}
\section{Trivial to quadrupolar phase transition}\label{sec_quad_phase transition}

	  We start by analysing the transition from the trivial to the quadrupolar phase, path I in Fig.~\ref{fig_phase}. The derivatives of the different contributions to the grand potential are shown in Fig.~\ref{fig_corr_g}. For these simulations, we take the intercell hopping parameters to be $\lambda_x=\lambda_y=1$ in a lattice of $40\times40$ unit cells and we use interpolation of the correlation functions, such that the higher derivatives can be calculated.   

	In Fig.~\ref{fig_corr_g}(a), we see that the first derivative of $\omega$ diverges for the corner contribution, while to spot the divergences of the grand potential for the edge and bulk we have to go to the second (Fig.~\ref{fig_corr_g}(b)) and third derivative (Fig.~\ref{fig_corr_g}(c)), respectively. This indicates that we can use the thermodynamic approach to identify this TQPT: the bulk exhibits a third-order phase transition, the edge a second-order, and the corner a first-order one. The main point is that the presence of the SPMs signals a discontinuity in the corresponding co-dimension, which is not captured when only considering the bulk free energy. These results are supported by a finite-size scaling shown in the Appendix A~\footnote{Notice that the edge and corner contributions are still finite, even in the trivial phase. These contributions are related to finite-size effects - which often come into play when considering non-extensive quantities - and go to zero when $\gamma$ is increased in the trivial phase.}.
	
	We can further understand this behavior by analysing the critical exponents using Josephson's hyperscaling relation.

\subsection{Josephson hyperscaling relation}\label{subsec_josephson}
    For quantum phase transitions, the canonical critical exponent $\alpha$ determines how the grand potential scales with the reduced phase transition parameter $t$,
	\begin{equation}
		\Omega\left(t_{-}\right)-\Omega\left(t_{+}\right)\propto \left|t\right|^{2-\alpha}, 
	\end{equation}
where $t_{+}$($_{-}$) indicates that we are approaching $t=0$ from positive (negative) values of $t$. An $n$-order derivative of $\Omega$ will scale with $t^{2-\alpha-n}$. Since t $\rightarrow 0$  at the phase transition, the derivatives will start to diverge for large enough values of $n$. This implies that $2-\alpha$ is the order of the phase transition because a derivative of order higher than $2-\alpha$ will diverge.

	We can determine this critical exponent, and therefore the order of the phase transition using the Josephson hyperscaling relation~\cite{continentino1994quantum, *continentino2017quantum}
	\begin{equation}
		2-\alpha=\nu\left(d+z\right),
		\label{eq_josephson}
	\end{equation}
where $d$ is the dimension of the system and the critical exponents $z$ and $\nu$ are determined by the way how the band gap $\Delta$ closes,
	\begin{equation}
	\begin{cases}
		\Delta\left(\mathbf{p},t=0\right)\propto \left|\mathbf{p}\right|^z\\
		\Delta\left(\mathbf{p}=0,t\right)\propto \left|t\right|^{\nu z}.
	\end{cases}
		\label{eq_crit_gap}
	\end{equation}
Here, $p=k-K$ is the momentum relative to the point $K$, at which the gap closes.

    For fully-periodic boundary conditions, $\lambda_x=\lambda_y=1$ and $\gamma_x=\gamma_y=\gamma$, the dispersion relation reads~\cite{benalcazar2017_Science,*benalcazar17_PRB}
    \begin{align}
	\begin{split}
		\varepsilon_{\pm}\left(\mathbf{k}\right)=&\pm \sqrt{2+2\gamma^2+2\gamma\left[\cos\left(k_x\right)+\cos\left(k_y\right)\right]}.
	\end{split}
		\label{eq_disp_gen}
	\end{align}
	
	The reduced phase transition parameter takes the form $t=\left(\gamma_c-\gamma\right)/\gamma_c$, where $\gamma_c$ is the value of the tuning parameter at which the gap closes. For this dispersion, the gap closing occurs at the $\Gamma$ point when $\gamma_c=-1$ and at $\mathbf{K}_{+}=\left(\pm\pi,\pm\pi\right)$ when $\gamma_c=1$. Then, we can obtain the behavior of the gap near the phase transition and determine the critical exponents using Eq.~\eqref{eq_crit_gap}. 
	
	For $\gamma_c=1$ and $\mathbf{k}=\mathbf{K}_{+}+\mathbf{p}$,
	\begin{align}
	\begin{split}
		\Delta\left(\mathbf{p},\gamma=1\right)\approx&2\sqrt{2}\left|\mathbf{p}\right|,
	\end{split}
	\end{align}
	which implies that $z=1$ for this transition.
	\begin{figure}[!ht]
		\centering
		\includegraphics[width=\linewidth]{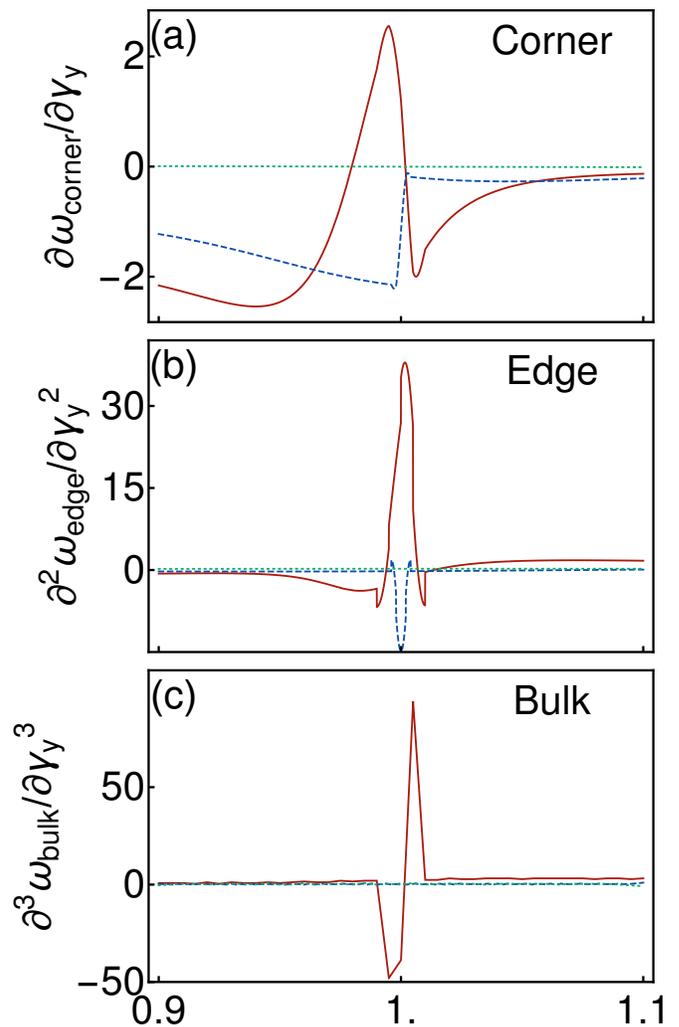}
		\caption{(Color online) (a) First derivative of the corner contribution, (b) second derivative of the edge contribution and (c) third derivative of bulk contribution to the grand potential with respect to $\gamma_y$. We consider the following three transitions: trivial to quadrupolar ($\gamma_x=\gamma_y=\gamma$, path I in Fig.~\ref{fig_phase}, red solid line), dipolar to quadrupolar ($\gamma_x=0.5$, path II in Fig.~\ref{fig_phase}, blue dashed line) and trivial to dipolar ($\gamma_x=1.5$, path III in Fig.~\ref{fig_phase}, green dotted line). The lattices have $40\times 40$ unit cells.}
		\label{fig_corr_gx}
	\end{figure}
	\begin{figure*}[!ht]
		\centering
		\includegraphics[width=\linewidth]{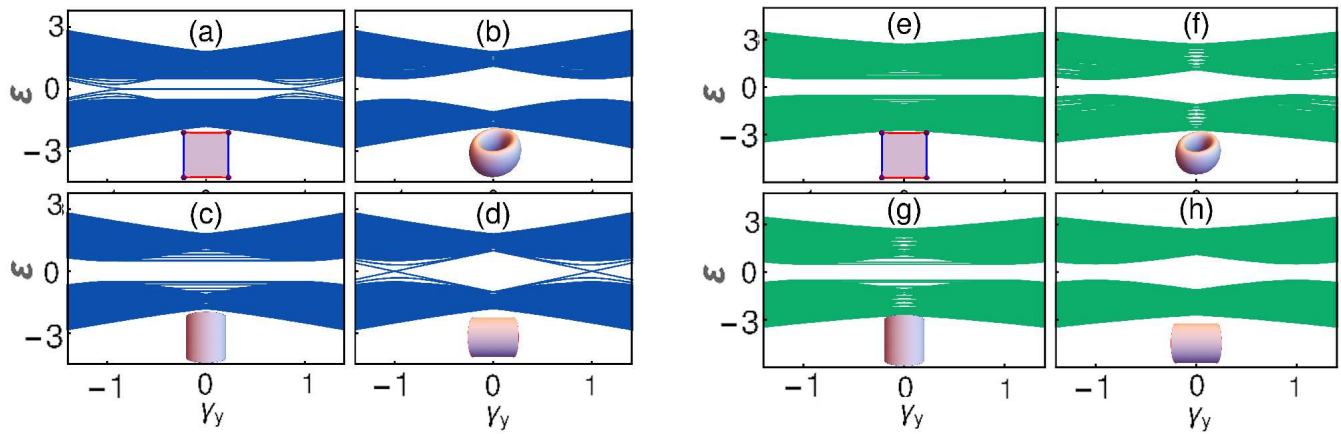}
		\caption{(Color online) Spectrum as function of $\gamma_y$ for $\gamma_x=0.5$ (path II in Fig.~\ref{fig_phase}) when the system has (a) open boundary conditions, (b) fully-periodic boundary conditions, (c) $x$-periodic boundary conditions and (d) $y$-periodic boundary conditions and for $\gamma_x=1.5$ (path III in Fig.~\ref{fig_phase}) when the system has (e) open boundary conditions, (f) fully-periodic boundary conditions, (g) $x$-periodic boundary conditions and (h) $y$-periodic boundary conditions. In comparison with Fig.~\ref{fig_spectrum}, we observe that the spectra for $x$-periodic and $y$-periodic boundary conditions are now different due to the absence of $C_4$ symmetry. The lattices have $20\times 20$ unit cells.}
		\label{fig_spectrum_gx}
	\end{figure*}
	For $\mathbf{k}=\mathbf{K}_{+}$ and $\gamma=\left(1-t\right)$,
	\begin{align}
	\begin{split}
		\Delta\left(\mathbf{k}=\pm \pi,t\right)=&2\sqrt{2}\left|t\right|,
	\end{split}
	\end{align}
	such that $\nu=1$ for this transition. The results are similar for $\gamma_c=-1$.
	
	Then, the Josephson hyperscaling relation implies that
	\begin{equation}
		2-\alpha=d+1.
		\label{eq_crit_d}
	\end{equation}
    The factor $2-\alpha$ denotes how the grand potential scales with $\gamma$ at the phase transition. Therefore, this relation predicts that the transition should be third-order for the bulk ($d=2$), which was also obtained from the correlation function. In order to interpret our numerical results, we associate the phase transition in the bulk - related to the bulk-band gap closing - with the one at the boundary - the appearance/disappearance of the SPM. If we assume that they have the same critical exponents $z$ and $\nu$, the only difference in the order of the phase transition should be attributed to the different dimension of each contribution. This reasoning works well for systems that exhibit a bulk-boundary correspondence \cite{kempkes2016universalities, van2018thermodynamic}, but it breaks down when the correspondence is not present \cite{cats2018staircase}. Using now that the bulk is 2D, the edge 1D and the corner 0D in Eq.~\eqref{eq_crit_d}, we find that the corner has a first-order phase transition and the edge a second-order one, as we obtained from the correlation-function method.

\section{Fixed $\gamma_x$ phase transitions}\label{sec_dip_phase transition}

	We now apply the same method to investigate two other phase transitions that can be tuned in this system: the one between the dipolar and the quadrupolar phase (obtained by fixing $\gamma_x=0.5$ and varying $\gamma_y$, path II in Fig.~\ref{fig_phase}) and the transition between the trivial and the dipolar phase (obtained by fixing $\gamma_x=1.5$ and varying $\gamma_y$, path III in Fig.~\ref{fig_phase}). 

	For fixed $\gamma_x$, the parameter that tunes the phase transition is $\gamma_y$. Thus, instead of deriving the Hamiltonian~\eqref{eq_ham} with respect to both $\gamma_x$ and $\gamma_y$ like in Eq.~\eqref{eq_Omega_dgamma}, we derive it only with respect to $\gamma_y$ 
	\begin{align}
	\begin{split}
		\frac{\partial \Omega}{\partial \gamma_y}=\sum\limits_{\mathbf{R}}\Braket{c^\dagger_{\mathbf{R},1}c_{\mathbf{R},4}-c^\dagger_{\mathbf{R},2}c_{\mathbf{R},3}+h.c.}.
	\end{split}
		\label{eq_Omega_dgamma_y}
	\end{align}
	We can obtain the different contributions by replacing this correlation function in Eqs.~\eqref{eq_omega_bulk_dgamma},~\eqref{eq_omega_edge_dgamma} and~\eqref{eq_omega_corner_dgamma}. In Fig \ref{fig_corr_gx}, we plot the singular derivatives of each contribution, namely, we plot the first derivative of the corner contribution, the second one for the edge and the third one for the bulk for each of the different phase transitions, in order to compare them.
	
	Inspection of Figs.~\ref{fig_corr_gx}(a) and \ref{fig_corr_gx}(b) indicates that the correlation functions exhibits discontinuities at both edge and corner for the dipolar to quadrupolar phase transition (blue), with about half the amplitude of the discontinuities of the transition between the trivial and quadrupolar states. Also, there is no signal of phase transition for any component in the trivial to dipole transition (green dotted line) as well as for the bulk component in the dipolar to quadrupolar phase transition (blue dashed line).
	
	To better understand why there is no response in the grand potential for these transitions, we analyse how the spectrum of the system changes along these phase transitions. In Fig.~\ref{fig_spectrum_gx}, we display the spectrum for $\gamma_x=0.5$ (a-d) and $\gamma_x=1.5$ (e-h), while we vary $\gamma_y$. 
    \begin{figure*}[!ht]
		\centering
		\includegraphics[width=\linewidth]{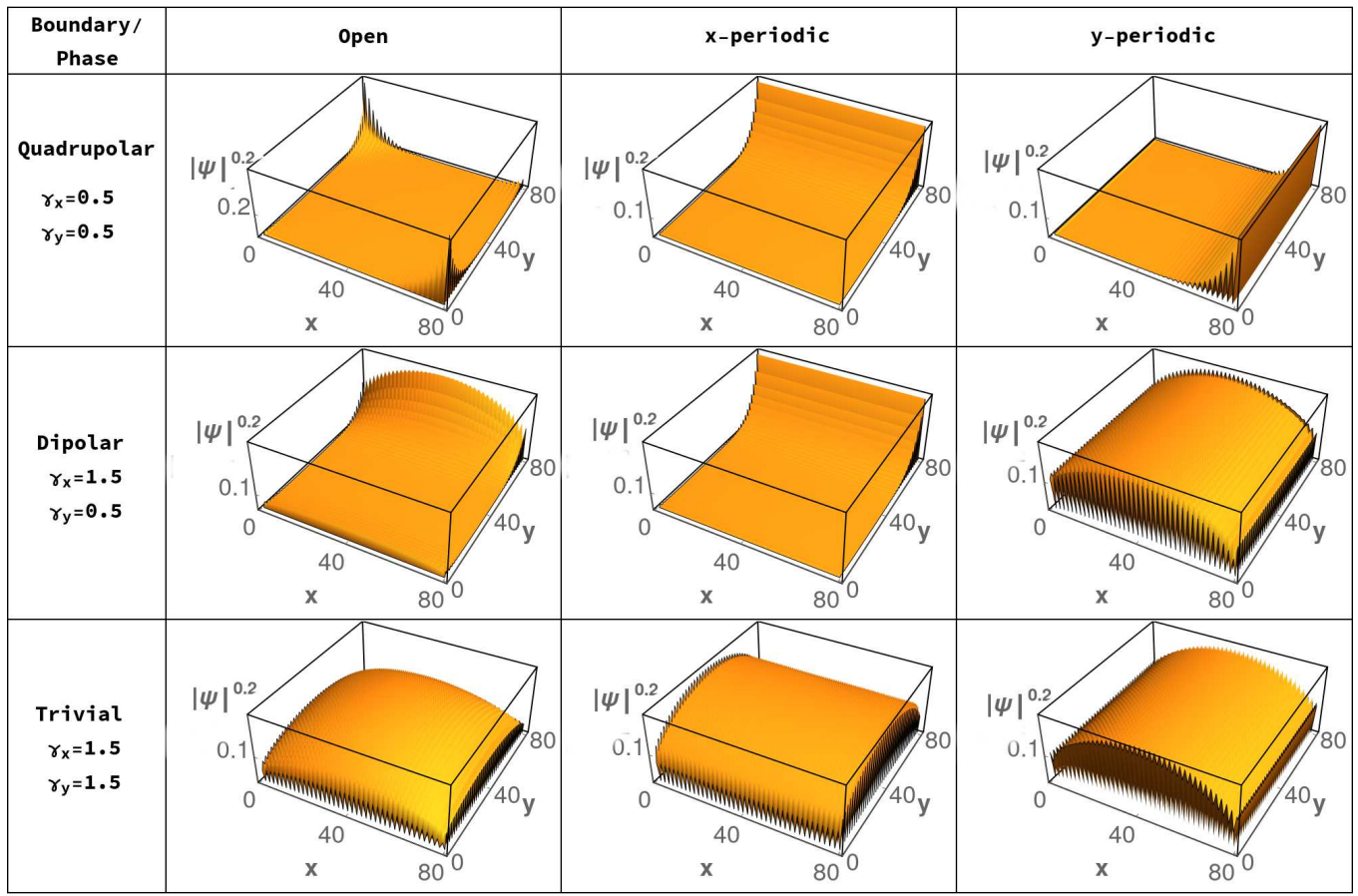}
		\caption{(Color online) Wavefunction for the symmetry-protected modes for several boundary conditions and for the different phases of the phase diagram shown in Fig.~\ref{fig_phase}. To better visualise the localisation of the wavefunction we actually plot $\left|\psi\right|^{0.2}$. These wavefunctions are evaluated for lattices with $40\times40$ unit cells.}
		\label{fig_wf}
	\end{figure*}
	For the dipolar to quadrupolar transition, there is a gap closing only at the $x$-edge (Fig.~\ref{fig_spectrum_gx}(d)) but not at the $y$-edge (Fig.~\ref{fig_spectrum_gx}(c)), whereas the bulk (which has periodic boundary conditions in both directions) does not present a gap closing (Fig \ref{fig_spectrum_gx}(b)). This seems to explain why there is no signature of phase transition at the bulk while the edge and corner contributions show discontinuities with nearly half the amplitude of the ones in the trivial to quadrupolar phase transition. For $\gamma_x=1.5$, none of the spectra shows a gap closing (Fig.~\ref{fig_spectrum_gx}(e)-(f)). Although there is no gap closing, the wavefunctions change across the phases (see Fig.~\ref{fig_wf}, where we plot the wavefunctions of the highest energy level in the valence band for each phase). We observe that for the dipolar phase with $\gamma_x=1.5$ and $\gamma_y=0.5$, the wavefunction for the $x$-periodic boundary conditions is identical to the one for the same boundary condition in the quadrupolar phase, so its localised in one of the $y$-edges and delocalised along $x$, see Fig.~\ref{fig_wf}. However, for $y$-periodic boundary conditions, the wavefunction is similar to the one in the trivial phase, which is delocalised, see Fig.~\ref{fig_wf}. These features are captured in the polarisation.
	
	The polarisation invariant changes across this phase transition without a gap closing because the symmetry that protects the polarisation is related to inversion symmetry, while the gap closing occurs only for systems with $C_4$ symmetry \cite{khalaf2019boundary}. The change in the polarisation is associated to a nested Wilson loop~\cite{benalcazar2017_Science, *benalcazar17_PRB}. However, the grand potential is not sensitive to the change in this invariant, as it is not sensitive to the presence or absence of inversion symmetry.
	
	The same occurs for the phase transition from the trivial to the dipolar phase. In this case, the corner, edge and bulk contributions are all flat (green lines in Fig.~\ref{fig_corr_gx}) and do not exhibit any sign of a phase transition. This behavior should be contrasted to the red curves in Fig.~\ref{fig_corr_gx}, which perfectly describe the TQPT from the trivial to the quadrupolar phases.
	
	This observation can be clarified in the context of the Josephson hyperscaling relation: it is impossible to define the critical exponents $z$ and $\nu$ because there is no actual gap closing at this phase transition. This has the quite striking implication that we are confronted with a TQPT that does not satisfy even Ehrenfest's classification, with all thermodynamical quantities varying smoothly across the phase transition. This kind of behavior is also seen in other types of phase transitions, like the Berezinskii-Kosterlitz-Thouless~\cite{kosterlitz1973ordering} one, which has an infinite order.
	
	\begin{figure}[!ht]
		\centering
		\includegraphics[width=\linewidth]{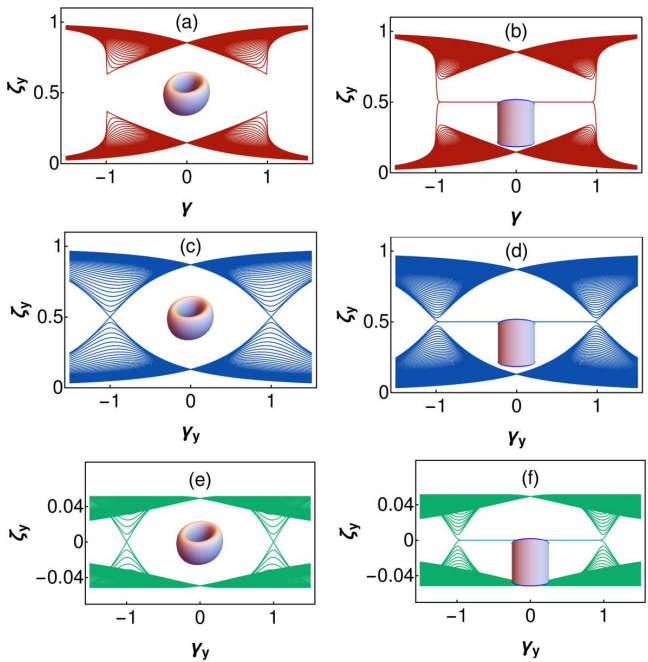}
		\caption{(Color online) Wannier spectrum for the trivial to quadrupolar phase transition (path I in Fig.~\ref{fig_phase}) for (a) fully-periodic boundary conditions and (b) $x$-periodic boundary conditions; for the dipolar to quadrupolar phase transition (path II in Fig.~\ref{fig_phase}) for (c) fully-periodic boundary conditions and (d) $x$-periodic boundary conditions; and for the trivial to dipolar phase transition (path III in Fig.~\ref{fig_phase}) for (e) fully-periodic boundary conditions and (f) $x$-periodic boundary conditions. We see a signature of the phase transition in all the plots. We use a system with $100$ unit cells in the non-periodic direction and $100$ k-points to calculate the Wilson loops. Note that in Fig. (a-d) we choose a different gauge in relation to the one used in Fig. (e-d) to make more evident the occurrence of a phase transition.}
		\label{fig_wan_spectrum}
	\end{figure}
	
	\section{Wannier grand potential}\label{sec_wann_free}
	
	    To extract thermodynamical information for this type of phase transitions, one should look at some quantities that do change across the different phase transitions. These are the Wannier centers, eigenvalues of the large Wilson loop. Consider the eigenstates of a (semi-) periodic system $\ket{\psi^{m}}\left(\mathbf{k}\right)$ with band index $m$ and labelled by the momentum $\mathbf{k}$ in the periodic direction. There is a matrix $\mathbb{G}\left(\mathbf{k}\right)$ associated with these eigenstates, which has the matrix elements
	   \begin{equation}
	        \mathbb{G}^{mn}\left(\mathbf{k}\right)=\Braket{\psi^{m}\left(\mathbf{k}+\Delta\mathbf{k}\right)|\psi^{n}\left(\mathbf{k}\right)},
	   \end{equation}
where $m$ and $n$ are occupied bands and $\left|\Delta \mathbf{k}\right|=2\pi/L$. $\mathbb{G}$ is called the Wilson loop element \footnote{The actual Wilson loop element will be the unitary version of this matrix, as this matrix is unitary only in the thermodynamical limit, as discussed in~\cite{benalcazar2017_Science, *benalcazar17_PRB}.}. We can then define the large Wilson loop as the successive multiplication of these elements
	   \begin{equation}
	        \mathbb{W}_{\mathcal{C},\mathbf{k}}=\mathbb{G}\left(\mathbf{k}-\Delta\mathbf{k}\right)\cdots\mathbb{G}\left(\mathbf{k}+\Delta\mathbf{k}\right)\mathbb{G}\left(\mathbf{k}\right),
	   \end{equation}
where $\mathcal{C}$ denotes a closed path in the first Brillouin zone that goes from $\mathbf{k}$ to $\mathbf{k}$. $\mathbb{W}$ is a unitary matrix, such that its eigenvalues are of the form $e^{\imath 2\pi \zeta}$; due to topology~\cite{benalcazar2017_Science, *benalcazar17_PRB, neupert2018lecture}, these eigenvalues always come in complex conjugated pairs. As a consequence, the Wannier centers $\zeta$ always come in pairs of positive and negative values. The Wilson loop can be written in terms of an effective Hamiltonian, the so called Wannier Hamiltonian $H_{\text{WC}}$
	   \begin{equation}
	        \mathbb{W}_{\mathcal{C},\mathbf{k}}=e^{\imath 2\pi \left[H_{\text{WC}}\left(\mathbf{k}\right)+\mu\right]},
	        \label{eq_wan_ham}
	   \end{equation}
such that its eigenvalues are the Wannier centers (at least $\mod 1$) and yield the Wannier spectrum. There is a gauge degree of freedom in the definition of this Hamiltonian, as we can add a term proportional to the identity to it without changing the Wilson loop. This term can be interpreted as an effective Wannier chemical potential \cite{khalaf2019boundary}.
	 
	   \begin{figure*}[!ht]
		\centering
		\includegraphics[width=\linewidth]{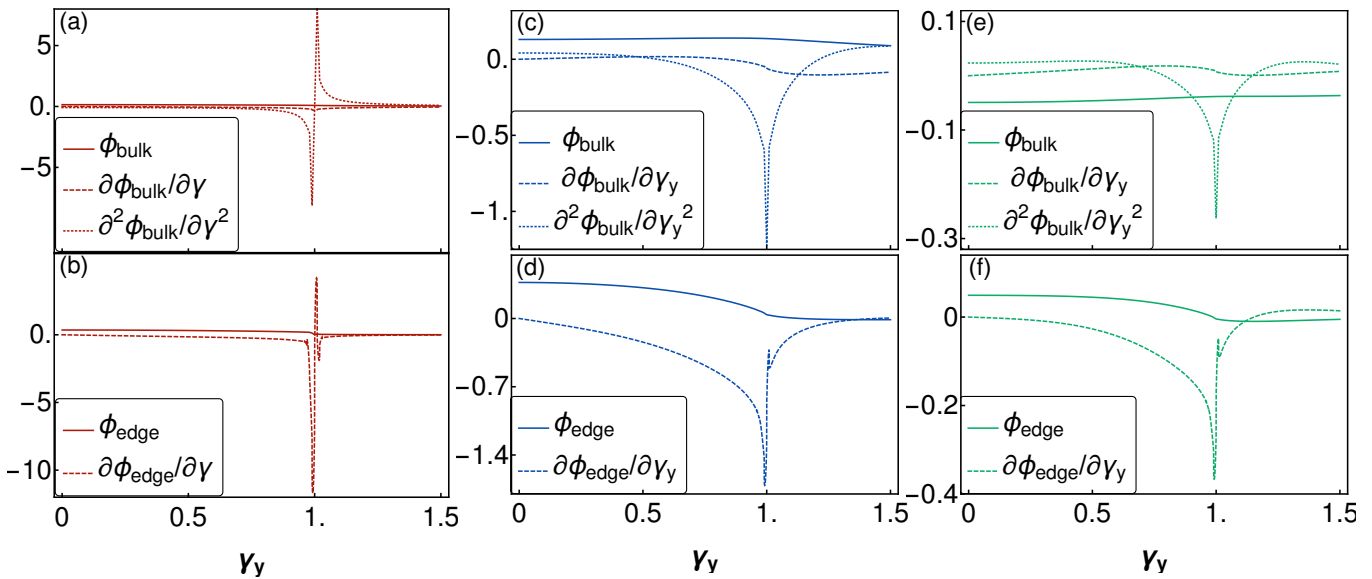}
		\caption{(Color online) Wannier grand potential (solid lines) and its first (dashed lines) and second (dotted lines) derivatives for the (a) bulk and (b) edge contributions along the trivial to quadrupolar phase transition (path I in Fig~\ref{fig_phase}, in red). (c) Bulk and (d) edge contributions along the dipolar to quadrupolar phase transition (path II in Fig~\ref{fig_phase}, in blue). (e) Bulk and (f) edge contributions along the trivial to dipolar phase transition (path III in Fig~\ref{fig_phase}, in green). We use a system with $200$ unit cells in the non-periodic direction and $100$ k-points to calculate the Wilson loops.}
		\label{fig_wan_free}
	\end{figure*}

    One can, in principle, use this effective Hamiltonian to obtain these Wannier bands and derive an effective grand potential from it, the Wannier grand potential $\Phi_W$. To test whether one can see a gap closing or any signature of the phase transitions analysed here, we show in Fig.~\ref{fig_wan_spectrum} the Wannier spectrum for this transition both for the fully-periodic and $x$-periodic Hamiltonian, considering a path $\mathcal{C}$ that goes from $k_x=-\pi$ to $k_x=\pi$. There are clear signatures in the Wannier spectrum of all the phase transitions, either a gap closing (or appearance of "horns" in the trivial to quadrupolar phase transition) for the fully-periodic boundary condition, or the appearance of in-gap states, which are related to a nontrivial polarization, for $x$-periodic boundary conditions. 
	    Since the exact form of the Wannier Hamiltonian in terms of operators is unknown, we are unable to identify which correlation functions are related to the derivative of $\Phi_W$. However, as argued in Refs.~\cite{benalcazar2017_Science, *benalcazar17_PRB, fidkowski2011model, khalaf2019boundary}, the Wannier Hamiltonian of Eq.~\eqref{eq_wan_ham} is adiabatically connected to the edge Hamiltonian with the Wannier chemical potential being related to the filling of the edge levels. In this way, the thermodynamics extracted from a Wannier Hamiltonian should correspond to the thermodynamics of a system with a 1D bulk and a 0D boundary.  In this case, we assume a form for $\Phi_W$ that is composed of a bulk $\phi_{\text{bulk}}$ and edge $\phi_{\text{edge}}$ contribution as
	    \begin{equation}
	        \Phi_W=\phi_{\text{bulk}}L+\phi_{\text{edge}},
	        \label{eq_wan_free}
	    \end{equation}
such that we can obtain both contributions by
	    \begin{equation}
	        \phi_{\text{bulk}}=\frac{\Phi_W^{\text{per}}}{L}
	        \label{eq_wan_free_bulk}
	    \end{equation}
and
	    \begin{equation}
	        \phi_{\text{edge}}=\Phi_W^{\text{x-per}}-\Phi_W^{\text{per}}.
	        \label{eq_wan_free_edge}
	    \end{equation} 

	    For free fermions at zero temperature, the grand potential defined in Eq.~\eqref{eq_gran_pot} is given by
	    \begin{align}
		\begin{split}
			\Omega=&-\lim\limits_{T\rightarrow 0}\frac{1}{\beta}\sum\limits_{m}\ln \left[1+e^{-\beta \left(\varepsilon_{m}-\mu\right)}\right]\\
			=&\sum\limits_{\varepsilon_m\leq \mu}\left(\varepsilon_m-\mu\right),
			\label{eq_Omega}
		\end{split}
		\end{align} 
where $m$ are all available states.

	    In an analogous way, the Wannier grand potential at $T=0$ is defined by
	    \begin{equation}
            \Phi_W=\sum\limits_{\zeta\leq \mu}\zeta,
            \label{eq_phi_nu}
        \end{equation}
where now the chemical potential $\mu$ should be set to $0.5$ for the trivial to quadrupolar or dipolar to quadrupolar phase transition, as is evident in the spectra of Fig.~\ref{fig_wan_spectrum}. This definition at zero temperature will be associated to the polarisation, with the difference that the Wannier grand potential is uniquely defined by the filling of the edge Hamiltonian. Note that Eq.~\eqref{eq_Omega} provides a more general description of the system at finite temperatures if the limit $T\rightarrow 0$ is not taken.
	    
	    We calculate these quantities using the Wannier spectra and the results are presented in Fig.~\ref{fig_wan_free}. The phase transitions for the 1D bulk are second order, while for the 0D edge these transitions are first order. It is interesting to notice that even in the absence of an actual gap closing in the trivial to quadrupolar phase transition, the Wannier grand potential is sensitive to the phase transition.   
	
	     As the Wannier gap closes with the same critical exponents as the energy gap of Eq.~\eqref{eq_crit_gap} (see Appendix B), these two transitions belong to the same universality class. We see that in this case, these phase transitions indeed satisfy Josephson's hyperscaling relation given in Eq.~\eqref{eq_crit_d} for a 1D bulk and a 0D edge. The hyperscaling relation together with the scaling of the Wannier grand potential reinforces the identification of the Wannier Hamiltonian with the edge Hamiltonian. 
	     
	     The fact that we can identify all the phase transitions using the Wannier grand potential indicates that the Wilson loop carries topological information that is not contained in the usual band structure, which is already suggested by the absence of any gap closing or appearance of in-gap states in the spectra of Fig.~\ref{fig_spectrum_gx}. However, assuming that the Wannier centers are indeed adiabatically connected to the edge spectrum, it is quite surprising that they yield a signature of the phase transition, while the actual spectrum, that contains the edge levels, does not provide this information. A possible answer to this question is that, as pointed out in Ref.~\cite{fidkowski2011model}, the density matrix obtained from the Wannier Hamiltonian is basically a purification of the density matrix obtained from the physical Hamiltonian. In this way, it might be possible to obtain the Wannier grand potential after a partial trace in the grand potential over the degrees of freedom related to the periodic direction that compose the large Wilson loop, but this is beyond the scope of this work.  	
\section{Conclusions}\label{sec_conc}

	    We extended the formalism used in Refs.~\cite{quelle2016thermodynamic, kempkes2016universalities, van2018thermodynamic, cats2018staircase,yunt2019topological} to investigate HOTIs, where the bulk-boundary correspondence relates the closure of the band gap with the zero modes that occur at the corners of the system. We numerically calculated the spectrum of Hamiltonian~\eqref{eq_ham}, which was proposed in Ref.~\cite{benalcazar2017_Science}, to identify the discontinuities in the derivative of the grand potential and elucidate the order of the TQPT. We found that for the trivial to quadrupolar phase transition, the bulk exhibits a third-order, the edge a second-order and the corner a first-order phase transition, in agreement with Josephson's hyperscaling relation.	
     
	    However, some of the TQPTs are not captured by the thermodynamic potential description, since those showed no discontinuities at transitions between states with different polarisations. This happens due to the absence of a band gap closing at these phase transitions. We then constructed a Wannier grand potential using the Wannier centers obtained via the Wilson loop, as the Wannier bands are expected to track the TQPT in situations when there is no energy gap closing.
	    
	    The Wannier grand potential presents critical exponents that have, according to Josephson hyperscaling relation, the same universality class of a system with a 1D bulk and a 0D edge. This, together with the scaling of the Wannier grand potential, confirms the identification of the Wannier Hamiltonian with the edge Hamiltonian.

	    A deeper understanding of this Wannier grand potential and its implication for evaluations of the entanglement entropy or Otto heat cycles, as well as a more complete comprehension of how one can relate the actual grand potential to the Wannier grand potential are interesting topics for future work. 

\section*{Acknowledgements}

	We thank Carlo Beenakker and Tom O' Brien for the access to the Maris cluster in Leiden, where some of the simulations were performed. We are grateful to Jette van den Broeke and Peter Cats for fruitful discussions. We would like to thank Raquel Queiroz and Sander Kooi for bringing  Ref.~\cite{khalaf2019boundary} to our attention. This work was mainly developed during the period when RA was a PhD student visitor at Utrecht University via the Delta Institute for Theoretical Physics (Delta ITP) consortium, a program of the Netherlands Organization for Scientific Research (NWO) that is funded by the Dutch Ministry of Education, Culture and Science. RA acknowledges funding from the Brazilian National Council for Scientific and Technological Development (CNPq) and the Brazilian Coordination for the Improvement of Higher Education Personnel (CAPES). SNK is financed by NWO under grant 16PR3245.
\clearpage

\begin{widetext}
	\begin{figure*}
		\centering
		\includegraphics[width=\linewidth]{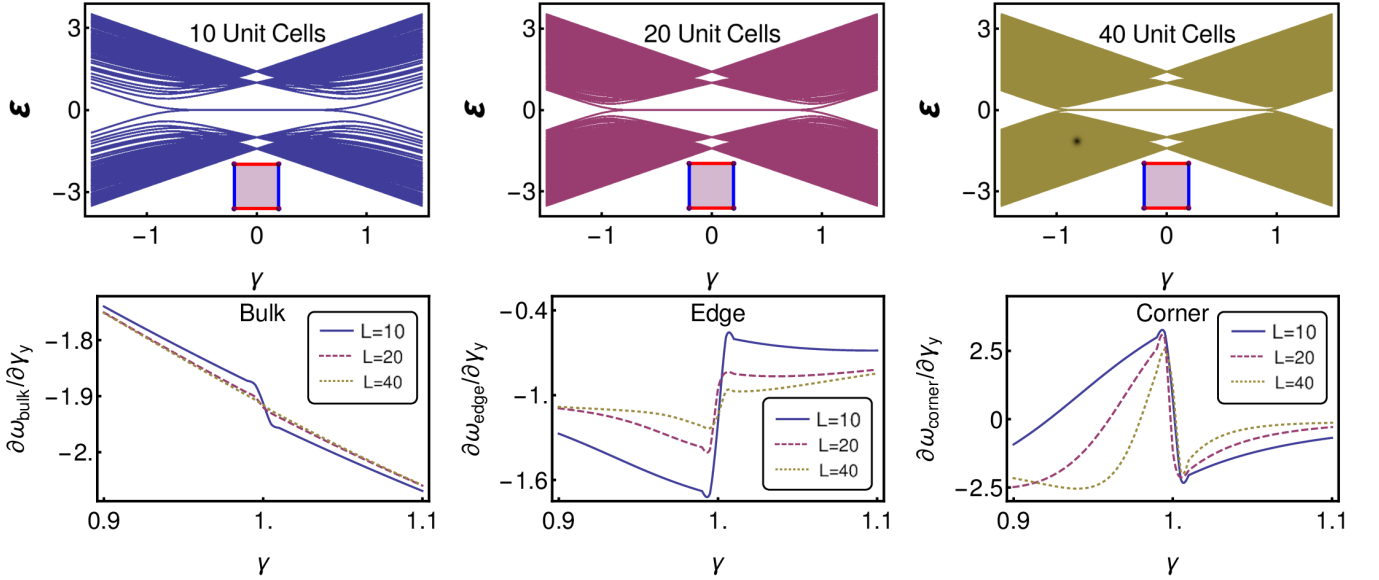}\\
		\caption{(Color online) Finite size effects. Spectrum and potential contributions for systems with $10\times10$ (blue solid lines), $20\times20$ (purple dashed lines) and $40\times40$ (yellow dotted lines) unit cells for the trivial to quadrupolar phase transition (path I in Fig.~\ref{fig_phase}).}
		\label{fig_finite_size}
	\end{figure*}
\end{widetext}
\appendix
\section{Finite size effects}\label{app_finite}

Here we comment on the main problem that arises due to finite-size effects, that is the hybridisation between corner modes when the system size is too small and the gap does not close at $\left|\gamma\right|=1$. In this case, the corner modes hybridize and move away from zero, while the bulk gap remains open.  As the correlation function is calculated from the spectrum, we can use the gap closing as a guideline to verify whether the system is large enough to reveal the phase transition.
 In Fig.~\ref{fig_finite_size}, we show the finite-size effects for the first derivative of the bulk, edge and corner contribution for different systems sizes. Although a small kink appears close to $\gamma=1$ for the bulk and edge first derivative, it becomes smaller and broader when we increase the system size, indicating a finite-size effect. In comparison, the divergence of the corner contribution becomes narrower as we increase the system size, indicating that is indeed related to a discontinuity.\\
\newpage
\section{Behavior of the Wannier gap}
\begin{figure}[!htb]
		\centering
		\includegraphics[width=\linewidth]{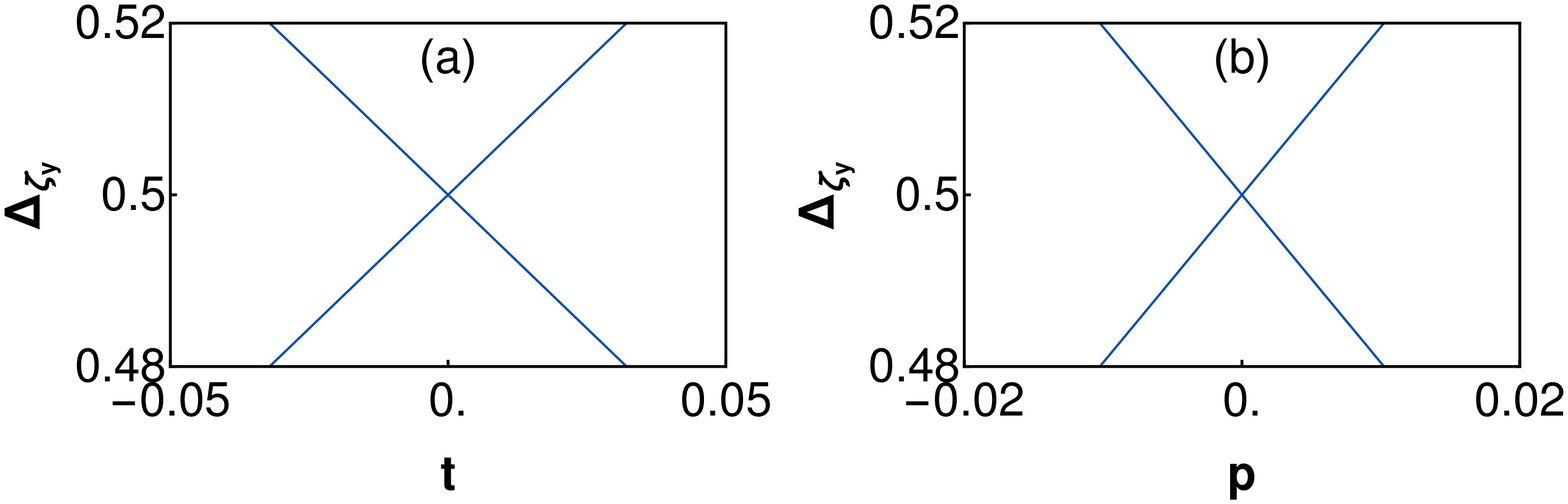}\\
		\caption{(Color online) The Wannier gap closes linearly with both (a) $t=\gamma_y-1$ ($k_x=\pi$) and (b) $p=\left(k_x-\pi\right)/\pi$ ($\gamma_y=1.0$).}
		\label{fig_wan_gap_closing}
	\end{figure}

Here, we show that the Wannier gap also closes linearly both in $\mathbf{p}$ and $t$, such that $\nu=1$ and $z=1$, as commented in Section \ref{sec_wann_free}. In Fig.~\ref{fig_wan_gap_closing}, we show that the Wannier gap closes linearly at the TQPT for both $t=\gamma_y-1$ ($k_x=\pi$) and $p=\left(k_x-\pi\right)/\pi$ ($\gamma_y=1.0$) for the quadrupolar to dipolar phase transition. The other phase transitions show the same critical exponents. 

\bibliography{hoti}

\end{document}